\documentclass{article}
\usepackage{longtable}
 \textwidth 15 cm
\title{On a classification of mesons }
\author{Sze Kui Ng\\
{\small Department of Mathematics, Hong Kong Baptist University,
Hong Kong}}
\begin{document}
\maketitle
\begin{abstract}
We give a mass formula for computing the mass spectrum of mesons. By this formula we show that  there are many mesons with their masses  corresponding to a prime number. In particular we show that all strange mesons are with their masses  corresponding to a prime number. With these prime numbers indexing the mesons we give a classification of mesons. We set up a knot model of mesons to derive this mass formula. In this knot model   mesons and their anti-particles are modeled by knots and their mirror images respectively. Then the amphichiral knots which are equivalent to their mirror images are used to model mesons which are identical with their anti-particles. With this knot model we show that there is a periodic phenomenon in the classification of mesons such that the starting nonet and the ending nonet are nonets of pseudoscalar mesons
with the $\pi$ meson modeled by an amphichiral knot. From this periodic phenomenon we give a theoretical argument for the existence of charm-anticharm mesons.
\end{abstract}

We give a formula to compute the mass spectrum of mesons, as follows \cite{Ng1}. Let $M$ denote the mass of a meson. Then we propose the following formula for $M$:
\begin{equation}
M= a\times b Mev
\label{1}
\end{equation}
where $a$ and $b$ are integers such that $a$ is as a consecutive number for indexing the meson; $b$ is as a number related to the total angular momentum $J$ of the meson. This number $b$ is proportional to the winding number of a knot model of the meson while $a$ is as an index of the knot model of the meson \cite{Ng1}. The number $b$ is around a number which is determined to be the number $45$. As examples let us consider the nonet of pseudoscalar mesons. By using (\ref{1}) we have that $\pi(135)$ is with the mass $3\times 45=135 Mev$ where $a=3$ is as a consecutive number and $b=45$ is as a number proportional to the winding number of a knot model of the meson $\pi(135)$. The number $45$ is related to the total angular momentum $J$ of $\pi(135)$. Similarly by using (\ref{1}) we have that the masses of $\eta(549)$ and $\eta^{'}(958)$ are given by $13\times 42= 546 Mev$ and $23\times 42=966 Mev$ respectively where $a=13$ and $a=23$ are as  indexes for the knot models of $\eta(549)$ and $\eta^{'}(958)$ respectively.
Let us then consider strange mesons. For strange mesons let us modify (\ref{1}) by introducing a number for the strange degree of freedom. We consider the following mass formula of a strange meson:
\begin{equation}
M= a\times b + n\times s Mev
\label{2}
\end{equation}
where the number $s$ equals $12$ or $24$ is for the strange degree of freedom of the strange meson; $n$ is an integer as the excited level of state of the strange meson and is also related to the total angular momentum $J$ of the strange meson. As an example let us consider the strange meson $K(498)=K^0(498)$.
By using (\ref{2}) we have that the mass of $K(498)$ is given by $11\times 43 +1\times 24= 497 Mev$ where $a=11$, $b=43$, $n=1$ and $s=24$. The two numbers $43$ and $1$ approximately correspond to that $J=0$ for $K(498)$.
Here the number $n=1$ gives the excited level of the strange state of $K(498)$.
Let us then consider mesons of the form $s\overline{s}$ where $s$ denotes the strange quark. For such mesons let us modify (\ref{1}) by introducing a strange-antistrange degree of freedom by the following formula:
\begin{equation}
M= a\times b + n\times s\overline{s} Mev
\label{3}
\end{equation}
where  $s\overline{s}$ denotes a number equal to $16$ or $32$ which is for the strange-antistrange degree of freedom of mesons of the form $s\overline{s}$ and $n$ is an integer as the excited level of strange-antistrange state of the $s\overline{s}$ meson and is also related to the total angular momentum $J$ of the $s\overline{s}$  meson. As an example let us consider the $\phi(1020)$ meson which is of the form $s\overline{s}$. By using (\ref{2}) we have that the mass of $ \phi(1020)$ is given by $23\times 43 +1\times 32= 1021 Mev$ where $a=23$, $b=43$, $n=1$ and $s\overline{s}=32$. The number $43$ and $1$ approximately correspond to that $J=1$ for $\phi(1020)$. Here we have that $32=2\times 16$ and we use the multiples of $16$ as the $s\overline{s}$ degree of freedom. We remark that this $s\overline{s}$ degree of freedom is a multiple of $16$ and is not a multiple of $12$ (or $24$) where $12$ (or $24$) is for the strange degree of freedom. We shall explain this phenomenon in a more complete investigation of this knot model of mesons \cite{Ng1}.
With the above three formulas we then compute the mass of the light mesons listed in \cite{PDA}\cite{PDA2}. We list the results of this computation in the following table which is as a classification of mesons.
\begin{longtable}{|c|c|c|c|}\hline $
 I=1 $&$ I=0
 $&$ \mbox{Strange mesons}\quad I=\frac12$&$ I=0
$ \\ \hline $
 \pi(135) $&$ \eta(549)$&$ K(498)$&$ \eta^{\prime}(958)$ \\ $
{\bf 3}\times 45=135$&$ {\bf 13}\times 42=546 $&$ {\bf 11}\times 43+24=497$&${\bf  23}\times 42=966  $ \\ \hline $

\rho(770)$&$ \omega(783)$&$K^{*}(896)  $&$\phi(1020) $ \\
$ {\bf 17} \times 45=765$&${\bf 17}\times 46=782 $&${\bf 19}\times 46+24=898$&$
{\bf 23}\times 43+32=1021 $ \\ \hline $

b_1(1235)$&$h_1(1170) $&$K_1(1270) $&$h_1(1380) $ \\
$ 27 \times 45=1242$&$26\times 45=1170  $&${\bf 29 }\times 43+24=1271 $&$
30\times 46= 1380  $ \\ \hline $

a_1(1260)$&$f_1(1285) $&$K_1(1400) $&$f_1(1420) $ \\
$ 28 \times 45=1260$&$28 \times 46=1288 $&${\bf 31}\times 44+36=1400 $&$
{\bf 29}\times 49=1421 $ \\ \hline $

a_0(980)$&$f_0(980) $&$ K_0^*(1430)$&$f_0(1370) $ \\
 ${\bf 23}\times 43=989 $&${\bf 23}\times 43=989 $&${\bf 31}\times 45+36=1431 $&$
{\bf 29} \times 47=1363 $ \\ \hline $

a_2(1320)$&$f_2(1270) $&$K_2^*(1430) $&$f_2(1430) $ \\
 $30\times 44=1320$&${\bf 29} \times 44= 1276
 $&${\bf 31}\times 45+36=1431 $&${\bf 31}\times 46=1426  $ \\ \hline $

a_2(1320)$&$f_2(1270) $&$K_2^*(1430) $&$f_2^{'}(1525) $ \\
 ${\bf 31}\times 43=1333$&${\bf 31} \times 41= 1271
 $&${\bf 31}\times 45+36=1431 $&${\bf 37}\times 41=1517 $ \\ \hline $

\rho(1450)$&$\omega(1420) $&$K^*(1410) $&$\phi(1680) $ \\
${\bf 31}\times 47=1452 $&${\bf 31}\times 46=1426 $&${\bf 31}\times 45+12=1407  $&$
{\bf 37}\times 45+16=1681$ \\ \hline $

\pi(1300)$&$\eta(1295) $&$K(1460) $&$\eta(1440) $ \\
$ {\bf 31} \times 42=1302$&${\bf 31}\times 42=1302 $&${\bf 31}\times 46+36= 1462 $&$\eta_H(1475): {\bf 37}\times 40=1480$ \\
$ {\bf 29} \times 45=1305$&$ $&$ $&$\eta_L(1410):32\times 44=1408 $ \\ \hline $

X(1600)=\pi_2(1600) $&$f_2(1565) $&$K_2(1580) $&$ f_2(1640) $ \\
$ 34\times 47=1598 $&$ 34\times 46=1564 $&${\bf 37}\times 42+24=1578 $&${\bf 41}\times 40 =1640 $ \\ \hline $

\pi_1(1405)$&$f_1(1510) $&$K_1(1650) $&$X(1650) $ \\
$ 32\times 44=1408$&$32\times 47=1504$&${\bf 37}\times 44+24=1652 $&$
36\times 46=1656 $ \\ \hline $

\rho(1700)$&$\omega(1600) $&$K^*(1680) $&$X(1910)=\phi(1910) $ \\
$ {\bf 37}\times 46=1702 $&${\bf 37}\times 43=1591 $&${\bf 37}\times 45+12=1677 $&$
{\bf 41}\times 46+16=1902 $ \\ \hline $

\pi_2(1670)$&$\eta_2(1645) $&$K_2(1770) $&$\eta_2(1870) $ \\
$ {\bf 37}\times 45=1665 $&$35\times 47=1645 $&${\bf 37}\times 47+36=1775 $&$ 39\times 48=1872$ \\ \hline $

\rho_3(1690)$&$\omega_3(1670) $&$K_3^*(1780) $&$\phi_3(1850) $ \\
$ 36\times 47=1692$&${\bf 37}\times 45=1665 $&$ {\bf 37}\times 47 +36=1775$&$ {\bf 41}\times 44+48=1852
$ \\ \hline $

 X(1775)=\pi_2(1775)$&$f_2(1810) $&$ K_2(1820) $&$ f_2(1950) $ \\
$ {\bf 37}\times 48=1776$&$ {\bf 37}\times 49=1813$&$
 {\bf 37}\times 49+12=1825$&$ 39\times 50=1950$ \\ \hline $

\pi(1800)$&$\eta(1760) $&$K(1830) $&$\eta(2225)
 $ \\
$ {\bf 43}\times 42=1806 $&${\bf 43}\times 41=1763 $&${\bf 43}\times 42+24=1830 $&${\bf 53}\times 42= 2226 $ \\ \hline $

a_0(1450)$&$f_0(1500) $&$K_0^*(1950) $&$f_J(1710) $ \\
$ 33\times 44=1452$&$33\times 45=1495 $&${\bf 41}\times 44+24=1951  $&$ 38\times 45=1710$ \\ \hline $

X(2000)=a_2(2000) $&$ f_2(2010)$&$ K_2^*(1980) $&$ f_2(2300) $ \\
$ {\bf 41}\times 49=2009 $&${\bf 41}\times 49=2009
 $&${\bf 41}\times 48+12=1980 $&$ {\bf 47}\times 49=2303 $ \\ \hline $

a_4(2040)$&$f_4(2050) $&$K_4^*(2045) $&$f_4(2300) $ \\
$ 40\times 51=2040$&$ {\bf 41}\times 50=2050 $&${\bf 41}\times 49+36=2045 $&$ {\bf 47}\times 49=2303$ \\ \hline $

\pi_2(2100)$&$f_2(2150)$&$K_2(2250) $&$f_2(2340) $ \\
$ {\bf 43}\times 49=2107$&${\bf 43}\times 50=2150  $&${\bf 43}\times 52+12=2248 $&${\bf 47}\times 50=2350 $ \\ \hline $

a_0(2020)$&$f_0(2020) $&$ K_0^*? $&$f_0(2200)$ \\
$ {\bf 43}\times 47=2021$&${\bf 43}\times 47=2021 $&$ {\bf 47}\times 47+36=2245$&$ 44\times 50=2200 $ \\ \hline $

\rho(2150)$&$\omega(2145) $&$K^*?$&$X(2340)=\phi(2340)$ \\
$ {\bf 43}\times 50=2150 $&${\bf 43}\times 50=2150 $&$ {\bf 43}\times 52+12=2248$&${\bf 47}\times 49+32=2335 $ \\ \hline $

\rho_3(2250)$&$\omega_3(2250) $&$K_3(2320) $&$X(2680)=\phi_3(2680)? $ \\
$ {\bf 47}\times 48=2256$&${\bf 47}\times 48=2256 $&$ {\bf 47}\times 49+12=2315$&${\bf 53}\times 50+32=2682
 $ \\ \hline $

\rho_5(2350)$&$X(2440)=\omega_5(2440) $&$K_5^*(2380) $&$X(2680)=\phi_5(2680) $ \\
$ {\bf 47}\times 50=2350$&${\bf 47}\times 52=2444  $&${\bf  47}\times 50+36=2386 $&$ {\bf 53}\times 50+32=2682$ \\ \hline $

\pi_4(2250)$&$f_J(2220) $&$K_4(2500) $&$X(2360)=f_4(2360) $ \\
$ {\bf 47}\times 48=2256 $&${\bf 47}\times 47=2209 $&${\bf 53}\times 47 +12=2503 $&$ 49\times 48=2352
$ \\ \hline $

a_6(2450)$&$f_6(2510) $&$K_6^*? $&$X(2710)=f_6(2710)$ \\
$ {\bf 47}\times 52 =2444$&$ 48\times 52=2496$&${\bf 47}\times 53+12=2503 $&${\bf 53}\times 51=2703 $ \\
\hline $

X(2710)=\pi(2710)$&$X(2750)=\eta(2750) $&$K(3100) $&$X(3250)=\eta(3250) $\\
$ {\bf 59}\times 46=2714$&${\bf 61}\times 45=2745 $&${\bf 67}\times 46+ 24=3106 $&${\bf 71}\times 46=3266  $\\ \hline
\end{longtable}
It is interesting to note that in this table all the strange mesons are indexed by a prime number $a$. Also many mesons are indexed by a prime number $a$. In this table the prime indexed number $a$ is in bold face. We use this index $a$ and the total angular momentum $J$ as the basic characteristics for the classification of mesons. Thus the index $a$ is together with the usual classification $J^{PC}$ to give a classification of mesons. However we shall relax the classification with the index $PC$ that we allow a classification of mesons with mixing $PC$. We use the number $b$ and the number $n$ to roughly determine the $J$ of the meson for classification.
Let us first consider the first and the second row of this table. These two nonets are classified by four prime numbers $3,11,13$ and $23$ and three consecutive prime numbers $17,19,23$ respectively. The number $23$ is jumped from the second nonet to the first nonet and this jumping is regarded as the characteristic of mesons with the notation $\prime$ such as the meson $\eta^{'}(958)$.
The number $46$ is greater than $43$. This difference of winding numbers gives an estimate of $J$ that if the first nonet is with $J=0$ then the second nonet is with $J=1$.
Then starting from the third row of the table the classification is simpler in that each row of mesons is classified by at most two consecutive prime numbers except the two rows of scalar mesons started with $a_0(980)$ and $a_0(1450)$ respectively; and the three rows started with $\pi(1800), X(2000)$ and $a_4(2040)$ respectively; and the last row. This last low is interesting in that it is again a nonet of pseudoscalar mesons which is similar to the first nonet of pseudoscalar mesons in that the $\pi$ mesons in these two nonets are modeled by an amphichiral knot. We shall later investigate this periodic phenomenon in more detail.
In the row of the scalar mesons $a_0(980), f_0(980), K_0^*(1430)$ and $f_0(1370)$ we classify scalar mesons by three consecutive prime numbers $23, 29$ and $31$. A characteristic of this classification of scalar mesons is that the index $31$ for the strange meson $K(1430)$ is greater than the index $29$ for $f_0(1370)$ and this is different from that for the pseudoscalar and vector mesons of the first and second rows of the table.
We remark that the properties and classification of scalar mesons have been studied in detail. Experiments show that the scalar mesons $a_0(980),f_0(980)$ have some property of mesons of the form $s\overline{s}$ \cite{PDA}-\cite{Kam}. Here we show that $a_0(980),f_0(980)$ and $\eta^{'}(958)$ are indexed by the same prime number $23$ as the $s\overline{s}$ meson $\phi(1020)$. This argrees with the experiments on $a_0(980),f_0(980)$. Then the scalar mesons $a_0(1450), f_0(1500), K_0^*(1950)$
and $f_J(1710)$ ($J$ is determined to be $0$) are with indexes $33, 41, 38$ which are in the region of the three consecutive prime numbers $31, 37$ and $41$ and we may regard these four mesons as classified by the three consecutive prime numbers $31, 37$ and $41$. We note that the index for $K_0^*(1950)$ is the prime number $41$ which is greater than the prime number $37$ (or $38$) and this agrees with the classification for the scalar mesons $a_0(980), f_0(980), K_0^*(1430)$ and $f_0(1370)$.
In this classification table except the first and the last row  we have that the indexes $a$ for the mesons of the same row and of the first  and the second columns are of the same number except in some cases where these two indexes are slightly different but are still in the region between two consecutive prime numbers. This is the basic characteristic of this classification of mesons. From the table we see that the largest such exceptional difference is the pair $\pi_2(1670)$ and $\eta_2(1645)$ which are indexed by $37$ and $35$ respectively. We note that the exceptional first and last rows are similar in that they are indexed by four prime numbers while all other rows are indexed by at most three consecutive prime numbers. This similarity of the first and the last row  shows that this is a periodic phenomenon of the classification table.
Let us then consider the three rows started with the $\pi$ mesons $\pi_1(1405), \pi_2(2100)$ and $\pi_4(2250)$. These three rows are with mixing $PC$.
Comparing to the first row started with the $\pi(135)$ meson and with four prime numbers we see that these three rows are compressed to have only two consecutive prime numbers and that the first two members of each of these three rows are of the same index. It is for this characteristic that these three rows are formed as classification of mesons with $\pi$ mesons as the $I=1$ meson. We shall later show that although these three mesons are denoted as a $\pi$ meson they are not a real $\pi$ meson as the $\pi(135)$ in the sense that the
$\pi(135)$ is modeled by an amphichiral knot while these three $\pi$ mesons are modeled by nonamphichiral knots. It is also for this reason that a mixing $PC$ classification is allowed in the formation of the three nonets for these three $\pi$ mesons.
Let us then consider the two rows started with the mesons $X(1600)$ and $X(1775)$ respectively. These two rows are similar to the three rows started with the $\pi$ mesons $\pi_1(1405), \pi_2(2100)$ and $\pi_4(2250)$. Thus we predict that $X(1600)$ is the meson $\pi_2(1600)$ and $X(1775)$ is the meson $\pi_2(1775)$.
Similarly for the row started with $\pi_4(2250)$ we predict that the meson $X(2360)$ is the meson $f_4(2360)$ and that the meson $f_J(2220)$ is the meson $f_4(2220)$. This $f_J(2220)$is with the computed mass ${\bf 47}\times 47=2209$. Here we have $b=47$ which is larger than $46$ but is smaller than $50$. From this rough estimate we have that the $J$ of this meson is roughly between $2$ and $4$. This agrees with experiments on the $J$ for this meson \cite{PDA}\cite{PDA2}. Here we put it to this row to determine that the $J$ of this meson is $4$.
For the row started with the vector meson $\rho(1700)$ we have the meson $X(1910)$. Here from our classification we predict that $X(1910)$ is the $\phi(1910)$. For the row started with the meson $X(2000)$ we predict that this meson is the $a_2(2000)$.
Let us then consider the nineth row of mesons containing the $\eta(1440)$ meson. Experiments show that there are two peaks $\eta_H(1475)$ and $\eta_L(1410)$ for this
$\eta(1440)$ \cite{Cic}-\cite{Bai}. Here from our computation we have two results on the mass of $\pi(1300)$: ${\bf 31} \times 42=1302 Mev$ and
${\bf 29} \times 45=1305 Mev$. Thus for the two peaks $\eta_H(1475)$ and $\eta_L(1410)$ we may have two classifications $\pi(1302), \eta(1295), K(1460), \eta_H(1475)$ and $\pi(1305), \eta(1295), K(1460), \eta_L(1410)$.
Similarly for the meson $f_2(1270)$ in the sixth and seventh rows we have two computational results: ${\bf 29} \times 44=1276 Mev$ and ${\bf 31} \times 41=1271 Mev$. We let these two results correspond to the two mesons $f_2(1430)$ and $f_2^{'}(1525)$. This then gives two classifications: $a_2(1320)$, $f_2(1270)$, $K_2^*(1430)$, $f_2(1430)$ and $a_2(1320)$, $f_2(1270)$, $K_2^*(1430)$, $f_2^{'}(1525)$.
We remark that this double phenomenon
may be due to the closeness of the two prime numbers $29$ and $31$. This closeness gives two close computations for the mass of $f_2(1270)$ and for the mass of $\pi(1300)$.
Physically this means that a meson with a state indexed by $31$ (or $29$) can be easily transited to form a state indexed by $29$ (or $31$).
Let us then consider the row started with $\rho(2150)$. For this row we predict the existence of a $K^*$ meson with mass approximately equal to ${\bf 43}\times 52+12=2248 Mev$. This $K^*$ is close to the $K_2(2250)$. Experiments show that there are various peaks in strange meson systems in the $2150-2260 Mev$ region \cite{PDA}\cite{PDA2}.
We thus predict that this $K^*$ is in this region.
Similarly for the row started with the meson $a_0(2020)$ we predict that there is a $K_0^*$ meson with mass approximately given by ${\bf 47}\times 47 +36=2245 Mev$. This $K_0^*$ is also close to the $K_2(2250)$. We thus predict that this $K_0^*$ is also in this $2150-2260 Mev$ region of strange meson systems.
Similarly we consider the row started with $a_6(2450)$. For this row we predict the existence of a $K_6^*$ meson with mass approximately equal to ${\bf 47}\times 53+12=2503 Mev$. We predict that this $K_6^*$ is close to the $K_4(2500)$.

Let us now consider the last row of the classification table. We show that this row gives a periodic phenomenon of classification of mesons in a sense as follows. First we have that this row is again classified with four prime numbers as the first row of the classification table (For this last row the four prime numbers are consecutive). Then let us show that why the region of the $X(2710)$ meson contains a $\pi$ meson. To this end let us model mesons as knots, as follows.
First let us index the prime knots by prime numbers by the following table \cite{Ng1}
\begin{displaymath}
\begin{array}{|c|c|c|c|c|c|c|c|c|c|c|} \hline
\mbox{prime knot} & {\bf 3_1}& {\bf 4_1}& {\bf 5_1} & {\bf 5_2}& {\bf 6_1}
&{\bf 6_2}&{\bf 6_3}&{\bf 7_1}&{\bf 7_2}&{\bf 7_3}\\ \hline
 \mbox{prime number}& &3 & 5& 7&11  &13 & 17&19&23&29 \\\hline
\mbox{prime knot}&{\bf 7_4} & {\bf 7_5}&{\bf 7_6} &{\bf 7_7} & {\bf 8_1}
&{\bf 8_2} &{\bf 8_3}&{\bf 8_4} &{\bf 8_5} & {\bf 8_6} \\ \hline
\mbox{prime integer}&31 &37&41 & 43&47  &53 &59 &61 & 67&71  \\ \hline
\end{array}
\end{displaymath}
where the prime knot ${\bf 3_1}$ is assigned with the number $1$. We have that the prime knots ${\bf 4_1},{\bf 6_2}$ and ${\bf 8_3}$ are assigned with the prime integers $3, 17$ and $59$ respectively. These three knots are called amphichiral knot in that they are equivalent to their mirror images. It follows that these three knots are suitable to model elementary particles which are identical with their anti-particles (For those elementary particles which are not identical with their anti-particles we then model them by nonamphichiral knots in such a way that their anti-particles are modeled by the mirror images of these nonamphichiral knots). Thus we have that it is suitable to model $\pi(135)$ with the prime knot ${\bf 4_1}$ and to model $\rho(770)$ and $\omega(783)$ with the prime knot ${\bf 6_2}$ since these mesons are identifcal with their anti-particles. Now we have that the $K(3100)$ meson is assigned with the prime number $67$ and since the prime number $59$ is assigned to the amphichiral knot ${\bf 8_3}$ we have that the mass region of  $X(2710)$ which contains $f_6(2710)$ \cite{Roz} should also contain a $\pi$ meson since $X(2710)$ can be indexed by $59$ with the computed mass ${\bf 59}\times 46=2714 Mev$. Thus we have $X(2710)=\pi(2710)$.
We remark that we may also use nonamphichiral knots to model elementary particles which are identical with their
anti-particles by using an average of the nonamphichiral knot with its mirror image. Here the amphichiral knots ${\bf 4_1},{\bf 6_2}$ and ${\bf 8_3}$ are special in that they are only for modeling mesons which are identical with their anti-particles.
Then for a meson such as the $\pi(135)$ meson which is modeled by the amphichiral knot ${\bf 4_1}$ we have that this meson should have special property which is only from an amphichiral knot. For the $\pi(135)$ meson this special property is that it is the starting meson of the family of mesons. Then when $X(2710)$ is identified as a special $\pi$ meson modeled by the amphichiral knot ${\bf 8_3}$ with the index number $59$ it is then
similar to the the starting $\pi(135)$ meson modeled by the amphichiral knots ${\bf 4_1}$.
This thus gives a periodic phenomenon of the classification of mesons.
It is then interesting to note from the classification table that the prime numbers $59$ and $61$ are not occupied with the $K$ mesons while the consecutive prime numbers starting from $29$ upto $53$ are all occupied with the $K$ mesons. This is similar to that the prime numbers $3, 5, 7$ are not occupied with the $K$ mesons. Let us explain this phenomenon as follows. We have that $59$ should not be occupied with the $K$ mesons since $59$ is assigned to the amphichiral knot ${\bf 8_3}$ and the $K$ mesons are not identified with their anti-particles. Then let us suppose that there is a $K$ meson which occupies the prime number $61$. Then this $K$ meson should not be in a nonet with the $I=1$ nonstrange meson assigned with the number $59$ because the amphichiral knot ${\bf 8_3}$ is for mesons such as the $\pi$ meson with $I=1$ having the property that the index number for the $K$ meson does not consecutively follow $59$ (This is similar to the amphichiral knot ${\bf 4_1}$ for the meson $\pi(135)$). Thus in this nonet the meson with $I=1$ is assigned with an index number less than $59$. However we see from the classification table that starting from the second nonet the difference between  the index number of meson with $I=1$  and the index number for the $K$ meson in the same nonet is not more than the difference between the two consecutive prime numbers where the larger one is the prime index number of the $K$ meson. Thus according to this rule there will have no $K$ mesons occupying the prime number $61$.
Now  since there are no $K$ mesons occupying the prime numbers $59$ and $61$ this gives a room for introducing a new degree of freedom. We have that ${\bf 61}\times 50 +48=3098 Mev$ where $48=2\times 24$ is similar to $2\times 16$ for the $\phi(1020)$ meson where $2\times 16$ is for the $s\overline{s}$ degree of freedom of the $\phi(1020)$ meson. This approximates the experimemtal mass $3096 Mev$ of the $J/\psi$ meson which is of the form  where $c$ denotes the charm quark. The number $48=2\times 24$ is then for the $c\overline{c}$ degree of freedom of $J/\psi$ and thus the prime integer $61$ is assigned to the $J/\psi$ meson.
Then for the prime number $59$ we have ${\bf 59}\times 50 +24=2974 Mev$. This approximates the
experimemtal mass $2979 Mev$ of the $\eta_c(2979)$ meson. Thus  $\eta_c(2979)$  is indexed by the prime number $59$.
The excited level of $c\overline{c}$ of $\eta_c(2979)$ is proportional to $24$ which is half of $48$ of the $c\overline{c}$-meson $J/\psi$. This agrees with the property of $\eta_c(2979)$ that it is partially a $c\overline{c}$-meson but is not a complete
$c\overline{c}$-meson such as the $J/\psi$ meson.
We note that since $\eta_c(2979)$ is indexed by the prime number $59$ we have that $\eta_c(2979)$ is modeled by the amphichiral knot ${\bf 8_3}$.
This agrees with the usual classification that  $\eta_c(2979)$ is put into the beginning pseudoscalar nonet
and it is as a periodic phenomenon
that both $\pi(135)$ and $\eta_c(2979)$ are amphichiral knots.
Let us then  consider more $c\overline{c}$-mesons in the following table \cite{PDA2}
\begin{displaymath}
\begin{array}{|c|c|c|c|} \hline
\mbox{$c\overline{c}$-meson} & \mbox{computed mass} &\mbox{$c\overline{c}$-meson} & \mbox{computed mass}\\ \hline
 \eta_c(1S)(2979)&{\bf 59}\times 50+24=2974 &
\eta_c(2S)(3594)&{\bf 79}\times 45+36=3591\\ \hline

 J/\psi(1S)(3096)&{\bf 61}\times 50+48=3098 &
 \psi(2S)(3686)&{\bf 73}\times 50+36=3686\\ \hline

 \chi_{c0}(1P)((3415)&{\bf 71}\times 48=3408 &
\psi(3770)&{\bf 83}\times 45+36=3771 \\ \hline

 \chi_{c1}(1P)((3510)&{\bf 73}\times 48=3504 &
\psi(3836)&{\bf 83}\times 45+2\cdot48= 3831\\ \hline

 h_c(1P)(3526)&{\bf 73}\times 48+24=3528 &
 \psi(4040)&{\bf 89}\times 45+36=4041 \\ \hline

\chi_{c2}(1P)((3556)&{\bf 79}\times 45=3555 &
 \psi(4160)&{\bf 83}\times 50+2\cdot36=4162\\ \hline

 & &\psi(4415)&{\bf 97}\times 45+ 48=4413 \\ \hline
\end{array}
\end{displaymath}
where $12, 24, 36$ and $48$ are for the $c\overline{c}$ degree of freedom.
From this table we see that the computed masses approximate the experimentl masses. We notice that the computed masses of $\chi$ mesons do not have the term of multiple of $12, 24, 36$ and $48$. This agrees with the property of $\chi$  that they are not a complete $c\overline{c}$ meson. Also we notice that there are many $c\overline{c}$-mesons in this mass region. Let us give a knot model argument to explain this phenomenon. Let us extend the table of indexing prime knots by prime integers with the following table:
\begin{displaymath}
\begin{array}{|c|c|c|c|c|c|c|c|c|c|c|c|c|} \hline
\mbox{prime knot} &{\bf 8_7}&{\bf 8_8}&{\bf 8_9} &{\bf 8_{10}}&{\bf 8_{11}}
&{\bf 8_{12}}&{\bf 8_{13}}&{\bf 8_{14}}&{\bf 8_{15}}&{\bf 8_{16}} &{\bf 8_{17}} &{\bf 8_{18}} \\ \hline
 \mbox{prime number}&73 &79 &83&89&97&101 &103&107&109&113
&127 &131 \\ \hline
\end{array}
\end{displaymath}
In this table we list prime knots with eight crossings. In knot theory we have the fact that the sets of prime knots with even crossings such as ${\bf 6_{(\cdot)}}, {\bf 8_{(\cdot)}},{\bf 10_{(\cdot)}}$ contain amphichiral knots while
the sets of prime knots with odd crossings such as ${\bf 7_{(\cdot)}}, {\bf 9_{(\cdot)}}$ contain no (or only a few) amphichiral knots. In this  table of prime knots with eight crossings we have
amphichiral knots ${\bf 8_3},{\bf 8_9},{\bf 8_{12}}, {\bf 8_{17}}, {\bf 8_{18}}$ while there are no amphichiral knots of the form ${\bf 7_{(\cdot)}}$ and  ${\bf 9_{(\cdot)}}$. Since these amphichiral knots are suitable for modeling mesons which are identical with their anti-particles such as the $\pi$ and $c\overline{c}$ mesons
this explains that why there is a family of $c\overline{c}$ mesons appearing in this mass region. Here we predict that there will have $c\overline{c}$ mesons modeled by the amphichiral knots ${\bf 8_{12}},{\bf 8_{17}},{\bf 8_{18}}$ which are indexed by the prime numbers $101, 127$ and $131$ respectively.

We remark that the $\sigma$ meson \cite{PDA2} has not been included in the above classification. We shall at elsewhere give a knot model of this meson to show its existence and properties.


\begin{thebibliography}{38}

\bibitem{Ng1}
S. K. Ng, hep-ph/0208098, hep-th/0209143, hep-th/0210024.
\bibitem{PDA}
C. Caso {\bf et al.}(Particle Data Group), The European Physical
Journal C{\bf 3} 1(1998).
\bibitem{PDA2}
K. Hagiwara {\bf et al.} (Particle Data Group), Phys. Rev.D {\bf
66},010001(2002).

\bibitem{Bev}
E. van Beveren {\bf et al.} Zeitschrift fuer Physik, {\bf C30} 615
(1986).

\bibitem{Bev2}
E. van Beveren and G. Rupp, Eur. Phys. J. {\bf C10} 468 (1999).

\bibitem{Bev3}
F.Kleefeld, E.van Beveren, G.Rupp and M.D.Scadron, Phys.Rev. {\bf
D66} 034007 (2002).

\bibitem{Del}
R. Delbourgo, D. Liu and M.D. Scadron, Phys. Lett. {\bf
B446}332(1999).

\bibitem{Abe}
A. Abele, Adomeit and Amsler,  Phys. Rev. D {\bf 57} 3860 (1998).
\bibitem{Abe2}
A. Abele, Adomeit and Amsler,  Phys. Lett. B {\bf 380} 453 (1996).
\bibitem{Abe3}
A. Abele, Adomeit and Amsler,  Phys. Lett. B {\bf 385} 425 (1996).
\bibitem{Tor}
Tornqvist, Phys. Rev. Lett. {\bf 49}, 624 (1982).
 \bibitem{Tor2}
Tornqvist, Z. Phys. Rev. C{\bf 68}, 647 (1995).
\bibitem{Tor3}
Tornqvist and Roos, Phys. Rev. Lett. {\bf 76}, 1575 (1996).
\bibitem{Jan}
Janssen, Pearce, Holinde and Speth, Phys. Rev.  D {\bf 52}, 2690
(1995).
\bibitem{Ams}
Amsler, Anisovich and Spanier,  Phys. Lett. B {\bf 333} 277
(1994).
\bibitem{Ams2}
Amsler, Anisovich and Spanier,  Phys. Lett. B {\bf 355} 425
(1995).
\bibitem{Ams3}
Amsler, Anisovich and Brose,  Phys. Lett. B {\bf 342} 433 (1995).
\bibitem{Isg}
Weinstein and Isgur, UTPT {\bf 89} 03 (1989).
\bibitem{Gra}
Grayer, Nucl. Phys. B {\bf 75} 189 (1974).
\bibitem{Ros}
Rosselet {\it et al.} Phys. Rev. D {\bf 15} 574 (1977).
\bibitem{Bec}
Becker, Nucl. Phys. B {\bf 151} 46 (1979).
\bibitem{Kam}
Kaminski, Phys. Rev. D {\bf 50} 3145 (1994).

\bibitem{Cic}
C.Cicalo {\it et al.} Phys. Lett. B{\bf 462} 453(1999).
\bibitem{Ber}
A. Bertin {\it et al.} Phys. Lett. B{\bf 400} 226(1997).
\bibitem{Ber2}
A. Bertin {\it et al.} Phys. Lett. B{\bf 361} 187(1995).
\bibitem{Ams}
C. Amsler {\it etal.} Phys. Lett. B{\bf 358} 389(1995).

\bibitem{Aug}
J.E. Augustin {\it etal.}Phys. Rev. D{\bf 46} 1951(1992).
\bibitem{Aug2}
J.E. Augustin {\it etal.}Phys. Rev. D{\bf 42} 10(1990).
\bibitem{}
M.G. Rath {\it et al.} Phys. Rev. D{\bf 40} 693(1989).

\bibitem{Bol}
T. Bolton {\it et al.} Phys. Rev. Lett. {\bf 69} 1328(1992).
\bibitem{Edw}
C. Edwards {\it et al.} Phys. Rev. Lett. {\bf 49} 259(1982).
\bibitem{Bai}
Z. Bai {\it et al.} Phys. Rev. Lett. {\bf 65} 2507(1990).

\bibitem{Sch}
D.L. Scharre {\it et al.} Phys. Lett. B{\bf 97} 329(1980).
\bibitem{Arm}
T.A. Armstrong {\it et al.} Nuc. Phys. B{\bf 227} 365(1983). {\it
etal.}
\bibitem{Bau}
M. Baubillier {\it etal.} Nuc. Phys. B{\bf 183} 1(1983).
\bibitem{Cle}
W.E. Cleland {\it etal.} Nuc. Phys. B{\bf 184} 1(1983).
\bibitem{Chl}
P.V. Chliapniov{\it etal.} Nuc. Phys. B{\bf 158} 253(1983).
\bibitem{Lis}
D. Lissauer {\it etal.} Nuc. Phys. B{\bf 18} 491(1983).
\bibitem{Roz}
M. Rozanska {\bf et al.}, Nuc. Phys. B {\bf 162}, 505(1980).
\bibitem{Den}
D.L. Denney {\bf et al.}, Phys. Rev. D{\bf 28}, 2726(1983).
\bibitem{Ale}
 A.N. Aleev {\bf et al.}, PAN {\bf 56}, 1358(1993).
\bibitem{Ani}
A.V. Anisovich {\bf et al.}, Phys. Rev.B{\bf 517}, 6(2001).

\bibitem{God}
S. Godfrey and J. Napolitano, Rev. Mod. Phys.{\bf 71},1411 (1999).
\bibitem{Ng}
 S.K. Ng,
 math-QA/0008103.
\bibitem{Lic}
W.B.R. Lickorish, {\it An Introduction to Knot Theory},
(Springer
1997).

\bibitem{Mur}
K. Murasugi, {\it Knot Theory and Its Applications}, (Birkhauser Verlag,
1997).
\bibitem{Lic}
W.B.R. Lickorish, {\it An Introduction to Knot Theory}, (Springer,
1997).
\bibitem{Rol}
D. Rolfsen. {\it Knots and Links}, (Publish or Perish, Inc. 1976).


\end{thebibliography}
 \end{document}